\documentclass[a4paper,leqno]{article}
\usepackage{latexsym,amssymb,amsmath}
\usepackage{color,amsfonts}
\usepackage[english]{babel}

\newtheorem{theorem}{Theorem}[section]
\newtheorem{lemma}[theorem]{Lemma}
\newtheorem{proposition}[theorem]{Proposition}

\newtheorem{definition}[theorem]{Definition}
\newtheorem{remark}[theorem]{Remark}
\newenvironment{proof}[1][Proof]{\textbf{#1.} }{\ \mathbb Rule{0.5em}{0.5em}}

\newcommand{\po}{{\mathbb P}}
\newcommand{\ep}{{\varepsilon}}
\begin{document}

\title{Invariant Gibbs measures of the energy for shell models of turbulence;
the inviscid and viscous cases}
\author{Hakima Bessaih
\footnote{University of Wyoming, Department of Mathematics, Dept. 3036, 1000
East University Avenue, Laramie WY 82071, United States, bessaih@uwyo.edu}
\and Benedetta Ferrario
\footnote{Universit\`a di Pavia, Dipartimento di Matematica, via
  Ferrata 1, 27100 Pavia, Italy,
benedetta.ferrario@unipv.it}}

\date{\today}
\maketitle

\begin{abstract}
Gaussian measures of Gibbsian type  are associated with
some shell model of 3D turbulence; they are  constructed by means of
the energy, a conserved quantity for the 3D inviscid and unforced
shell model.
We prove the existence of a unique global flow for a stochastic viscous shell
model and of  a global flow for the deterministic inviscid shell
model, with the property that
these Gibbs measures are invariant for these flows.
\end{abstract}

\section{Introduction}
The study of existence and uniqueness of solutions for
incompressible inviscid and viscous flows
with initial data in some "physically relevant" space is of great
interest. The most understood case is the 2D model, for which
existence and uniqueness of
classical and weak solutions for the viscous flow with initial data of
finite energy are due to  J. Leray and later to O. Ladyszenskaja,
J.-L. Lions and G. Prodi,
while for the inviscid flow
they are due to W. Wolibner and later to V. Judovich, T. Kato and
C. Bardos with more assumptions on the initial velocity.
However, the 3D motion is a more challenging problem;
for the viscous case, Leray's work gives
existence but not uniqueness of weak solutions
for initial data of finite
energy,
whereas with more restrictive assumptions on the initial data there exists
a unique local solution. For the 3D inviscid case, only local results
for the well posedness of weak solutions
are known.
We refer to \cite{MP94} where the authors, C. Marchioro and M. Pulvirenti provide a comprehensive introduction to a wide range of topics related to equations of inviscid and incompressible fluid flow.
The interested reader can find a quite recent account of all these results in \cite{L}.

Inside the analysis of the equations of hydrodynamics, statistical
solutions have been investigated.
In fact the individual solutions may give a detailed
and too complicated picture of the fluid, while one could be
interested  in the behavior of some
global quantity related to the fluid,
where the microscopic picture is replaced by the macroscopic one.
This is the statistical approach to turbulence.
From the mathematical point of view, we are interested
in distributions invariant for these flows.
Probability measures of Gibbsian type,
with Gibbs density expressed by means of invariants of the 2D motions
have been discussed in \cite{ac,arfhk,AHK,AHK1,AHK2,BF,BPP,CDG,c,dapdeb,deb,af}.
The construction of such invariants measures looks quite natural
since the 2D Euler equation has the nice property to admit
infinitely many first integrals, including the quadratic invariants
given by  the energy and the enstrophy.
In particular, in all the previous papers the Gibbs density
is constructed by means of the enstrophy.
No results of Gibbs invariant measures are known for the
3D equations of hydrodynamics.

In this paper, we consider some shell models in a very general form which
includes the SABRA and the GOY models.
These models are the most interesting and most popular examples
of simplified phenomenological models of turbulence.
This is because, although departing from reality,
they capture some essential statistical properties
and features of turbulent flows, like the energy and the enstrophy
cascade and the power law decay of the structure functions
in some range of wave numbers, the inertial range.
From the computational point of view, shell models are much
simpler to simulate than the Navier-Stokes equations due to the
fact that  we need a moderate number of degrees of freedom
to reach high Reynolds numbers
(see, e.g., \cite{sabra}
and references therein).
Indeed,
shell models of turbulence describe the evolution of
complex Fourier-like components of a scalar velocity field
denoted by $u_{n}$ and the associated wavenumbers  are denoted by $k_{n}$,
where the the discrete index $n$ is referred as the shell index.
The evolution of the infinite sequence $\{u_n\}_{n=-1}^\infty$ is given by
\begin{equation}\label{SHELL}
\dot u_n(t)+\nu k_n^2 u_n(t)+b_n(u(t),u(t))=f_n(t, u(t)),
\qquad n=1,2,\ldots
\end{equation}
with $u_{-1}=u_0=0$ and $u_n(t) \in \mathbb C$ for $n \ge 1$. Here
$\nu \ge 0$ and in analogy with Navier-Stokes equations $\nu$
represents a kinematic viscosity;
$k_n=k_0 \lambda^n$ ($\lambda>1$) and $f_n$ is a forcing term.
The exact form of $b_n(u,v)$ varies from one model to another.
However in all the various models,
it is assumed that $b_n(u,v)$ is chosen in such a way that
\begin{equation}\label{incomp_shell}
 \Re \displaystyle\sum_{n=1}^{\infty}b_n(u,v)\overline{v}_{n}=0,
\end{equation}
where $\Re$ denotes the real part and $\overline x$ 
 the complex conjugate of $x$.
Equation \eqref{incomp_shell}
implies a formal law of conservation of energy in the inviscid
($\nu=0$) and unforced form of \eqref{SHELL}.
These models have similar properties to 3D fluids.

In particular, we define the bilinear terms $b_n$  as
\[b_n(u,v)=i(a k_{n+1}\overline u_{n+1} \overline v_{n+2}+
            b k_{n}\overline u_{n-1} \overline v_{n+1}-
            a k_{n-1}\overline u_{n-1} \overline v_{n-2}-
            b k_{n-1}\overline u_{n-2} \overline v_{n-1})
\]
in the GOY model (see \cite{G,goy})
and by
\[
b_n(u,v)=-i(a k_{n+1}\overline u_{n+1}  v_{n+2}+
            b k_{n}\overline u_{n-1} v_{n+1}+
            a k_{n-1} u_{n-1} v_{n-2}+
            b k_{n-1} u_{n-2} v_{n-1})
\]
in the SABRA model (see \cite{sabra}).
\\
The two parameters $a, b$ are real numbers.

In the present paper, we consider particular Gaussian  measures of Gibbs type
and investigate  their role in the analysis of shell models. Basically,
these Gibbs measures $\mu^\nu$ are constructed by means of the energy, which is
an invariant of motion for the inviscid and unforced shell
model. Therefore, our aim is to show
that these measures are invariant for the inviscid shell
model as well as for a suitable stochastic viscous shell model.
The support of the measure $\mu^\nu$ is a Sobolev space of negative
exponent and the space of finite energy initial velocity is
negligible with respect to the measure $\mu^\nu$.  Thus, one looks for
a flow with initial data of infinite energy.

Our results are very similar to those proved for the 2D stochastic
Navier--Stokes
and 2D deterministic Euler equation with respect to the Gibbs measure of the
enstrophy (a conserved quantity for the 2D equation of hydrodynamics)
in a series of papers \cite{ac,arfhk,AHK,AHK1,AHK2,c,dapdeb,deb,af}.
However, our results hold for general shell models
for which only the energy is an invariant of
 motion and are therefore  approximation models for 3D hydrodynamics.

Let us describe the content of the paper in more details.  In Section \ref{S-FS}
we introduce the equations, the
Gibbs measure  $\mu^\nu$ and their basic properties; in particular,
we introduce the Ornstein-Uhlenbeck equation with a suitable
 noise and prove that the Gibbs measure  $\mu^\nu$
is its unique invariant measure.
In Section \ref{S-viscoso} we focus on the stochastic
viscous shell model, having  $\mu^\nu$ as invariant measure;
first, we prove that for  $\mu^\nu$-a.e. initial data
there exists a unique global solution, and then that there exists a
unique stationary process whose law at any fixed time is  $\mu^\nu$.
The last Section \ref{S-eul} deals with the inviscid shell model,
for which we prove that there exists a
stationary process solving it and  whose law at any fixed time is  $\mu^\nu$.

%SSSSSSSSSSSSSSSSSSSSSSSSSSSSSSSSSSS
\section{Functional setting}\label{S-FS}
Even if in \eqref{SHELL}  we considered the unknowns $u_n(t)\in
\mathbb C$, from now on we deal with the real part and the imaginary
part of $u_n$: $u_{n,1}=\Re u_n$ and $u_{n,2}=\Im u_n$.
As usual,
for $x=(x_1,x_2) \in \mathbb R^2$ we set $|x|^2= x_1^2+x_2^2$ and
$x\cdot y=x_1y_1+x_2y_2$ is the scalar product in $\mathbb R^2$.

\subsection{Spaces and operators}
For any $\alpha \in \mathbb R$ set
$$
H^{\alpha}=\{u=(u_1, u_2, \ldots) \in (\mathbb R^2)^\infty:
 \sum_{n=1}^\infty k_n^{2\alpha}|u_n|^2<\infty \}.
$$
This is a Hilbert space with scalar product
$\langle u,v\rangle_{H^\alpha}=\sum_{n=1}^\infty k_n^{2\alpha} u_n \cdot
v_n$. Denote by  $\|\cdot \|_{H^\alpha}$  its norm.
We have the continuous embedding
$$
 H^{\alpha_1}\subset H^{\alpha_2} \qquad \text{ if } \alpha_1>\alpha_2.
$$

Let $A$ be the linear unbounded operator in $H^0$ defined as
\[
 A: (u_1, u_2, \ldots) \mapsto (k_1^2 u_1,k_2^2 u_2,\ldots), \;
 D(A)=H^2.
\]
The fractional power operators $A^p$ are well defined for any
$p \in \mathbb R$:
\[
 A^p: H^{2p+\beta}\to H^\beta,
 (u_1, u_2, \ldots) \mapsto (k_1^{2p} u_1,k_2^{2p} u_2,\ldots)
\]
in any space (i.e. for any $\beta$).
For any $p<0$, $A^p$ is a trace class operator in $H^\beta$, since
$Tr (A^p) = \sum_n k_n^{2p}=k_0^{2p} \sum_n \lambda^{2p n}$ is finite if
and only if $p<0$;
therefore, the operator
$A^{p}$ ($p<0$) is compact and
Hilbert--Schmidt as a linear operator in $H^{2p+\beta}$.

Moreover, $A$ generates an analytic semigroup of contractions in $H^0$
and for any $p>0$ and $t>0$
\begin{equation} \label{semigr}
 \|A^p e^{-\nu A t}x\|_{H^0}\le  \frac {c_{p,\nu}}{t^{p}} \|x\|_{H^0},
\end{equation}
with $c_{p,\nu}=(\frac p{e\nu})^p$.

Set $B_n=(B_{n,1},B_{n,2})$,
where $B_{n,1}$ and $B_{n,2}$ are, respectively,  the real part and
the imaginary part of the $b_n$ given in the previous section. For
instance, in the SABRA model
\begin{align}
\begin{split}
&B_{1,1}(u,v)=ak_{2} [-u_{2,2}  v_{3,1} +u_{2,1}  v_{3,2}]
\\
&B_{1,2}(u,v)=-a k_{2} u_{2} \cdot v_{3}
\end{split}
&\\
\begin{split}
&B_{2,1}(u,v)=ak_{3} [-u_{3,2}  v_{4,1} +u_{3,1}  v_{4,2}]
           +b k_{2} [-u_{1,2} v_{3,1}+u_{1,1} v_{3,2}]
\\
&B_{2,2}(u,v)=-a k_{3} u_{3}\cdot  v_{4} -b k_{2}  u_{1} \cdot v_{3}
\end{split}
&\\
\intertext{and for $n >2$}
\begin{split}
&B_{n,1}(u,v)= \; ak_{n+1} [-u_{n+1,2}  v_{n+2,1} +u_{n+1,1}  v_{n+2,2}]
\\&\qquad\qquad\qquad   +b k_{n} [-u_{n-1,2} v_{n+1,1}+u_{n-1,1} v_{n+1,2}]
\\&\qquad\qquad\qquad           +a k_{n-1}[u_{n-1,2} v_{n-2,1}+u_{n-1,1} v_{n-2,2}]
\\&\qquad\qquad\qquad           +b k_{n-1}[u_{n-2,2} v_{n-1,1}+u_{n-2,1} v_{n-1,2}],
\end{split}
&\\
\begin{split}
&B_{n,2}(u,v)=-a k_{n+1} [u_{n+1,1}  v_{n+2,1} +u_{n+1,2}  v_{n+2,2}]
\\&\qquad\qquad\qquad    -b k_{n}  [u_{n-1,1} v_{n+1,1}+u_{n-1,2} v_{n+1,2}]
\\&\qquad\qquad\qquad    -a k_{n-1} [u_{n-1,1} v_{n-2,1}-u_{n-1,2} v_{n-2,2}]
\\&\qquad\qquad\qquad    -b k_{n-1} [u_{n-2,1} v_{n-1,1}-u_{n-2,2} v_{n-1,2}].
\end{split}&
\end{align}
Define the bilinear operator $B:({\mathbb R}^2)^\infty\times ({\mathbb
R}^2)^\infty\to ({\mathbb R}^2)^\infty$ as
\[
 B(u,v)=(B_1(u,v), B_2(u,v),\ldots).
\]
We have that $B$ is well defined  when its domain is
$H^1 \times H^0$ or $H^1 \times H^0$ (see  \cite{clt}), that is
$ B:  H^1 \times H^0\to H^0$  and
$ B:  H^1 \times H^0\to H^0$ are bounded operators. We extend the
result of \cite{clt} to more general spaces; this is very similar to
Proposition 1 of \cite{clt2}.
     \begin{lemma}\label{Bgenerale}
For any $\alpha_1, \alpha_2, \alpha_3 \in \mathbb R$
$$
 B:H^{\alpha_1}\times H^{\alpha_2} \to H^{-\alpha_3} \;
 \text{ with } \alpha_1+\alpha_2+\alpha_3=1
$$
and there exists a constant $c$ (depending on $a,b,\lambda$ and  the
$\alpha_j$'s) such that
\[
 \|B(u,v)\|_{H^{-\alpha_3}}\le c \|u\|_{H^{\alpha_1}} \|v\|_{H^{\alpha_2}}
 \qquad  \forall u \in H^{\alpha_1}, v \in H^{\alpha_2}.
\]
\end{lemma}
\proof The proof is similar to that of  Proposition 1 in \cite{clt}. We write
it for reader's convenience.

First,
\[
 \|B(u,v)\|_{H^{-\alpha_3}}=\sup_{\|z\|_{H^{\alpha_3}}\le 1}
  \Big| \sum_{n=1}^\infty B_n(u,v)\cdot z_n \Big|.
\]
Now we estimate the trilinear term. Looking at the espression of the
$B_n$'s
we have eight series $\sum_{n=1}^\infty$
to consider. We write the details for the first
one, working on the others in the same way.
\begin{equation}\begin{split}
 \sum_{n=1}^\infty &|a k_{n+1} u_{n+1,2}  v_{n+2,1} z_{n,1}|
\\&=
 \sum_{n=1}^\infty a k_0 \lambda^{n+1} |u_{n+1,2}  v_{n+2,1} z_{n,1}|
\\&= a \lambda^{1-\alpha_1-2\alpha_2}
 \sum_{n=1}^\infty   (k_0\lambda^{n+1})^{\alpha_1} |u_{n+1,2}|
                    (k_0\lambda^{n+2})^{\alpha_2} |v_{n+2,1}|
                    (k_0\lambda^{n})^{\alpha_3} |z_{n,1}|
\\&= a \lambda^{1-\alpha_1-2\alpha_2}
 \sum_{n=1}^\infty   k_{n+1}^{\alpha_1} |u_{n+1,2}| \;
                    k_{n+2}^{\alpha_2} |v_{n+2,1}| \;
                    k_{n}^{\alpha_3} |z_{n,1}|
\\&\le
 a \lambda^{1-\alpha_1-2\alpha_2}
 \sqrt{\sum_{n=1}^\infty k_{n+1}^{2\alpha_1} |u_{n+1,2}|^2}
 \sqrt{\sum_{n=1}^\infty k_{n+2}^{2\alpha_2} |v_{n+2,1}|^2}
 \sup_n (k_{n}^{\alpha_3} |z_{n,1}|)
\\&\le
 a \lambda^{1-\alpha_1-2\alpha_2}
 \|u\|_{H^{\alpha_1}} \|v\|_{H^{\alpha_2}} \|z\|_{H^{\alpha_3}}.
\end{split}\end{equation}
\hfill $\Box$

In particular, $B$ is a bounded operator in the following spaces:
\begin{equation}\label{stime}
B:H^{-\alpha}\times H^{-\alpha} \to H^{-2\alpha-1},
\end{equation}
\begin{equation}\label{stime0-alpha}
B:H^{-\alpha}\times H^{0} \to H^{-\alpha-1} , \;\;
B:H^{0}\times H^{-\alpha} \to H^{-\alpha-1}
\end{equation}
and
\begin{equation}\label{stime-5}
B:H^{-2-\alpha}\times H^{-2-\alpha} \to H^{-5-2\alpha}
\end{equation}
for  $\alpha\in \mathbb R$.

A remarkable property of the  operator $B$ is
\begin{equation}\label{Benerg}
  \sum_{n=1}^\infty B_n(u,v)\cdot v_n=0
 \end{equation}
whenever $u$ and $v$ give sense to the l.h.s..

%SSSSSSSSSSSSSSSSS
\subsection{Gibbs measure of the energy}\label{2.3}
For any $\nu>0$,
let us define the probability measure $\mu^\nu$ on $({\mathbb R}^2)^\infty$ as
$$
 \mu^\nu=\otimes_{n=1}^\infty \mu^\nu_n,
$$
where $\mu^\nu_n$ is the  Gaussian measure
on $\mathbb R^2$:
$$
 \mu^\nu_n(du_n)
 =\frac \nu{2\pi} e^{-\nu \frac 12 (u_{n,1}^2+u_{n,2}^2)}du_{n,1}du_{n,2}.
$$
Heuristically we have
$$
 \mu^\nu(du)=''\frac 1Z e^{-\nu E(u)}du'',
$$
where 
$E=\frac 12 \sum_{n=1}^\infty |u_n|^2$ is the energy and 
$Z$ is a normalization constant to make $\mu^\nu$
to be a probability measure.
This is the reason to call $\mu^\nu$ the Gibbs measure of the energy
with parameter $\nu$.

The support of the measure $\mu^\nu$ is bigger than the space $H^0$ of
finite energy.
Indeed, for any $c>0$ we have
$ \mu^\nu(\{x \in ({\mathbb R}^2)^\infty: \sup_n |x_n|< c  \})=0$;
since the space $ H^0$ is contained in the space of bounded sequences,
we have also that
\[
  \mu^\nu(H^0)=0.
\]
Moreover
\[
\begin{split}
 \int \|u\|^2_{H^{\alpha}} \mu^\nu(du)
&=
 \sum_{n=1}^\infty k_n^{2\alpha}\iint_{\mathbb R^2}
 (u_{n,1}^2+u_{n,2}^2) \frac \nu{2\pi}
  e^{-\nu \frac 12(u_{n,1}^2+u_{n,2}^2)}du_{n,1}du_{n,2}
\\
&=\frac 2 \nu \sum_{n=1}^\infty k_n^{2\alpha}=
  \frac 2 \nu k_0^{2\alpha} \sum_{n=1}^\infty \lambda^{2\alpha n}.
\end{split}
\]
This is finite if and only if $\alpha<0$.
Hence,
\[
  \mu^\nu(H^{\alpha})=1 \qquad \forall \alpha<0.
\]
Thus we set
$$
 \mathbb H=\cap_{\alpha<0}H^{\alpha}.
$$
$\mathbb H$ is a Fr\'echet space (see, e.g., \cite{rs1}) and
\[
  \mu^\nu(\mathbb H)=1.
\]

According to Kakutani's theorem (see \cite{kuo}),
the measures $\mu^{\nu_1}$ and $\mu^{\nu_2}$ are orthogonal (i.e.,
mutually singular)
for $\nu_1 \neq \nu_2$ positive.

For $p\ge 1$ we denote by $\mathcal L^p_{\mu^\nu}$
the space of Borelian  functions
$\phi:\mathbb H\to \mathbb R$ such that $\int |\phi|^p \ d\mu^\nu<\infty$.
We have $\mathcal L^p_{\mu^\nu}\subseteq\mathcal L^q_{\mu^\nu}$ for
$p\ge q$.

Let $FC^\infty_b$ be the space of infinitely differentiable cylindrical
functions  bounded and with bounded derivatives,
that is
\[ \exists k\in \mathbb N: \;
\phi=\phi(x_{i_1},x_{i_2},\ldots,x_{i_k}) \in
C^\infty_b\big((\mathbb R^2)^k;\mathbb R\big) .
\]
Analogously let
$FPol$ be the space of cylindrical polynomial functions.
Either $FC^\infty_b$ or $FPol$
is a dense subspace of $\mathcal L^p_{\mu^\nu}$ for $1\le p<\infty$.

An important property is the integrability
of $B$ with respect to the measure $\mu^\nu$.
\begin{proposition}\label{Bstaz}
$$
 \int\|B(x,x)\|_{H^{-1-\alpha}}^p \mu^\nu(dx)<\infty
$$
for any $p\in \mathbb N$ and $\alpha>0$.
\end{proposition}
\proof We write the proof for $p=2$ but it is the same for the other
values of $p$, since $\mu^\mu$ is Gaussian and the $B_n$'s are
second order polynomial. The details are given for the SABRA model, but the
result is true for all ''finite'' shell models.

We have
\[\begin{split}
\int |B_{n,1}(x,x)|&^2 \mu^\nu(dx)
=
 \int |a k_{n+1} [-x_{n+1,2}  x_{n+2,1} +x_{n+1,1}  x_{n+2,2}]
\\&           +b k_{n} [-x_{n-1,2} x_{n+1,1}+x_{n-1,1} x_{n+1,2}]
\\&           +(a+b) k_{n-1}[x_{n-1,2} x_{n-2,1}+x_{n-1,1} x_{n-2,2}]|^2 \mu^\nu(dx)
\\&
\le 2 \int \{a^2 k^2_{n+1} [x^2_{n+1,2}  x^2_{n+2,1} +x^2_{n+1,1}  x^2_{n+2,2}]
\\&       +b^2 k^2_{n} [x^2_{n-1,2} x^2_{n+1,1}+x^2_{n-1,1} x^2_{n+1,2}]
\\&       +(a+b)^2 k^2_{n-1}[x^2_{n-1,2} x^2_{n-2,1}+x^2_{n-1,1} x^2_{n-2,2}]\}\mu^\nu(dx)
\\& =\frac{16}{\nu^2} \{a^2 k^2_{n+1}+b^2 k^2_{n}+(a+b)^2 k^2_{n-1}\}
\\&= \frac{16}{\nu^2}k_0^2\{a^2\lambda^4+b^2\lambda^2+(a+b)^2\}\lambda^{2(n-1)}.
\end{split}
\]
Similarly we estimate $\int |B_{n,2}(x,x)|^2 \mu^\nu(dx)$.
Therefore
\[\begin{split}
 \int \|B(x,x)\|_{H^{-1-\alpha}}^2\mu^\nu(dx)
 &=
 \int \sum_{n=1}^\infty k_n^{2(-1-\alpha)}|B_n(x,x)|^2 \mu^\nu(dx)\\
 &\le c_{\nu,k_0,\lambda} (|a|^2+|b|^2) \sum_{n=1}^\infty \lambda^{-2n\alpha}
\end{split}
\]
which is finite if and only if $\alpha>0$. \hfill $\Box$

\medskip
We give a definition.
\begin{definition}
We say that a process $v=\{v_t\}_{t\ge 0}$ is a $\mu^\nu$-stationary process if
\\i)
$v$ is a stationary process;
\\ii)
the law of $v(t)$ is $\mu^\nu$ for any $t\ge 0$.
\end{definition}

Notice that, if $v$
is a $\mu^\nu$-stationary process defined on $(\Omega,\mathbb F,\po)$,
denoting by $\mathbb E$ the mathematical expectation we have
\begin{equation}\label{BpT}
\begin{split}
\mathbb E\left[ \int_0^T\|B(v(t),v(t))\|^p_{H^{-1-\alpha}}dt\right]
&=
 \int_0^T \mathbb E [\|B(v(t),v(t))\|^p_{H^{-1-\alpha}}]dt\\
&=
 T\int_{\mathbb H} \|B(x,x)\|^p_{H^{-1-\alpha}}\mu^\nu(dx)<\infty.
\end{split}
\end{equation}
Therefore, we have the following result.
\begin{proposition}\label{corB}
Let $v$
be a $\mu^\nu$-stationary process.
Then, for any $p\ge 1, \alpha<0$ we have that
$B(v,v) \in L^p(0,T;H^{-1-\alpha})$ $\po$-a.s..
\end{proposition}

%SSSSSSSSSSSSSSSSS
\subsection{The Wiener process}
Let $(\Omega,\mathbb F,\po)$ be a complete probability space, with expectation
denoted by $\mathbb E$.
Consider a a sequence $\{w_{n,j}\}_{n \in \mathbb N; j=1,2}$ of independent
standard 1-dimensional Brownian motions defined for all real $t$.
We say that $w$ is an $H^0$-cylindrical Wiener process if
$$
 w=\big((w_{1,1},w_{1,2}),(w_{2,1},w_{2,2}),(w_{3,1},w_{3,2}),\ldots).
$$
We set $ {\mathbb F}_{t}=\sigma \{w(s_2)-w(s_1), s_1 \leq s_2 \leq t \}$.

The paths of the process $w$
are ($\po$-a.s.) in $C^\beta([t_0,T];H^{-\alpha})$ for any
$-\infty<t_0<T<+\infty$, $0\le \beta <\frac 12$ and $\alpha>0$.
In fact, $w(t)$ has values in  $H^{-\alpha}$ for any $\alpha>0$ since
\[
 \mathbb E \Big[\|w(t)\|^2_{H^{-\alpha}}\Big]
  =\mathbb E\Big[\sum_{n=1}^\infty   k_n^{-2\alpha}|w_n(t)|^2\Big]
  =2 |t|\sum_{n=1}^\infty   k_n^{-2\alpha}
  =2 k_0 ^{-2\alpha} |t|\sum_{n=1}^\infty   \lambda^{-2\alpha n},
\]
and the latter series converges if and only if $\alpha>0$.
\\
Moreover, with similar argument we get from Kolmogorov criterium
(see, e.g.,  \cite{dpz} Theorem 3.3) that
\begin{equation}\label{holder}
\mathbb E[\|w\|^2_{C^\beta([t_0,T];H^{-\alpha})}]<\infty \quad \text{
  for any } \beta \in [0,\tfrac 12), \alpha>0.
\end{equation}

%SSSSSSSSSSSSSSSSS
\subsection{The equations}\label{S-equ}
Let us consider the stochastic viscous shell model
\begin{equation}\label{equu}
 du(t)+[\nu Au(t)+B(u(t),u(t))]dt = \sqrt{2 A} \ dw(t).
\end{equation}
As we shall see in the next section,
the covariance of the Wiener process has been chosen in such a way
that the measure $\mu^\nu$ is invariant for \eqref{equu}
(in a sense to be specified
later on); with this type of covariance we cannot
analyze equation \eqref{equu} with classical techniques, as done for
instance in \cite{bbbf}.

When there is no viscosity and  forcing term in \eqref{equu}, we get the
deterministic unforced and inviscid shell model
\begin{equation}\label{eul}
 \frac{du}{dt}(t)+B(u(t),u(t))=0.
\end{equation}

From property \eqref{Benerg} we have that
the energy $E(t)=\frac 12 \sum_{n=1}^\infty |u_n(t)|^2$
is an invariant of motion for \eqref{eul}, i.e. for any $t$
$$
 \dfrac {dE}{dt}(t)=\sum_{n=1}^\infty \dot u_n(t) \cdot u_n(t)
 = - \sum_{n=1}^\infty B_n(u(t),u(t)) \cdot u_n(t)  =0,
$$
whenever we consider a dynamics giving sense to the latter quantities.

We are interested also in the linear stochastic equation, i.e. the 
Ornstein--Uhlenbeck equation
\begin{equation}\label{equz}
 dz(t)+\nu Az(t)\ dt = \sqrt {2A} \ dw(t).
\end{equation}
For any time interval $[t_0,T]$, this equation has a unique strong
solution
\begin{equation}\label{sol-z}
 z(t)=e^{-\nu (t-t_0)A}z(t_0) +\int_{t_0}^t e^{-\nu (t-s)A} \sqrt {2A} \ dw(s).
\end{equation}
This is easy to prove for this linear
stochastic equation, which corresponds
to an infinite system of decoupled linear equations
($n \in \mathbb N$, $j=1,2$)
\begin{equation}\label{eqzn}
 dz_{n,j}(t)+\nu k_n^2 z_{n,j}(t) dt =\sqrt 2  k_n dw_{n,j}(t).
\end{equation}
We have that $z(t)$ takes values in $H^{-\alpha}$ ($\alpha>0$) $\po$-a.s.
if $z(t_0)\in H^{-\alpha}$. Indeed,
if $z(t_0)$ is in $H^{-\alpha}$ then $e^{-\nu (t -t_0)A}z(t_0)$ stays
in the same space.
And for the stochastic integral we have
\[\begin{split}
 \mathbb E[\| \int_{t_0}^t e^{-\nu (t-s)A} \sqrt {2A} \ dw(s)\|_{H^{-\alpha}}^2]
 &=
 \mathbb E\sum_{n=1}^\infty \sum_{j=1}^2 k_n^{-2\alpha}
     |\int_{t_0}^t e^{-\nu (t-s)k_n^2} \sqrt{2}k_n dw_{n,j}(s)|^2
\\&=
 \sum_{n,j}  k_n^{-2\alpha} \int_{t_0}^t e^{-2\nu (t-s)k_n^2} 2k_n^2  ds
\\&
 \leq \frac 2\nu \sum_{n=1}^\infty  k_n^{-2\alpha} =
       \frac 2\nu k_0^{-2\alpha} \sum_{n=1}^\infty  \lambda^{-2\alpha n}.
\end{split}\]
Moreover, the paths are a.s. continuous in time. In fact,
the continuity of the trajectories is easily obtained,
because $A$ is a diagonal operator commuting with the  covariance
operator of  the Wiener process $w$ (see \cite{dpz}
 Theorem 5.9). Further (see \cite{dpz} Remark 5.11) we have
\[
 \mathbb E \sup_{0\le t \le T}\|z(t)\|^p_{H^{-\alpha}}<\infty
\]
for any $p\ge 1$.

Finally, the stationary solution to \eqref{equz} can be represented as
\begin{equation}\label{zstaz}
  \zeta(t)=\int_{-\infty}^t e^{-\nu (t-s)A} \sqrt {2A} \ dw(s),
\end{equation}
and the law of $\zeta(t)$ is $\mu^\nu$ for any $t$.

%%%SSSSSSSSSSSSSSSSSSSSSSSSSSSSSSSSSSSSSs
\subsection{Invariance of the measure $\mu^\nu$}\label{S-inva}
Let us consider how the measure $\mu^\nu$ is related to the three
equations considered in the previous section. We present well known
properties, which are similar to those for the 2D Navier-Stokes
equation with respect to the Gibbs measure of the enstrophy
(see \cite{ac,af,AHK,AHK1,AHK2,arfhk}).

We start with the easy linear stochastic case \eqref{equz}.
Denote by $z_x(t)$
the unique strong solution of equation \eqref{equz} started at time
$t=0$
from  $x$ and evaluated at time $t>0$; this has been given in
\eqref{sol-z}.

We have that the measure $\mu^\nu$ is the unique
invariant measure of
equation \eqref{equz}, in the sense that
\begin{equation}
 \int \mathbb E[\phi(z_x(t))]\  \mu^\nu(dx)= \int \phi(x) \  \mu^\nu(dx)
\qquad \forall t\ge 0, \phi \in \mathcal L^2_{\mu^\nu}.
\end{equation}
Indeed, we
define the Markov semigroup $\{R_t\}_{t\ge 0}$  as
\begin{equation}\label{semR}
 (R_t \phi) (x)=\mathbb E[\phi(z_x(t))].
\end{equation}
Hence the invariance of the measure $\mu^\nu$ is
\begin{equation}\label{OUinvar}
 \int R_t\phi(x)\  \mu^\nu(dx)= \int \phi(x) \  \mu^\nu(dx)
\qquad \forall t\ge 0, \phi \in \mathcal L^2_{\mu^\nu}.
\end{equation}

Formally, we have $R_t=e^{-tQ}$, $Q$ being
the Ornstein--Uhlenbeck operator.
The concrete expression of the
generator is easily given on particular dense subspaces of
$\mathcal L^2_{\mu^\nu}$. For $\phi \in FC^\infty_b$
we have
\begin{equation}
 Q\phi(x)=\sum_n \sum_{j=1}^2\Big[
  k_n^2 \frac{\partial^2 \phi}{\partial x_{n,j}^2}(x)
    - \nu k_n^2 x_{n,j}\frac {\partial \phi}{\partial x_{n,j}}(x)\Big]
\end{equation}
Here for $\phi:(\mathbb R^2)^k\to \mathbb R$, $\phi\in FC_b^\infty$
we set
$$
 \tilde \phi(x_{i_1,1},x_{i_1,2},x_{i_2,1},x_{i_2,2},\ldots,x_{i_k,1},x_{i_k,2})=
 \phi(x_{i_1},x_{i_2},\ldots,x_{i_k})
$$
and
\[
 \frac{\partial \phi}{\partial x_{n,j}}(x_{i_1},x_{i_2},\ldots)=
 \frac{\partial \tilde \phi}{\partial x_{n,j}}
   (x_{i_1,1},x_{i_1,2},x_{i_2,1},x_{i_2,2},\ldots).
\]

Hence \eqref{OUinvar} is
equivalent to the infinitesimal invariance
\begin{equation}\label{Q-inv}
 \int Q\phi \ d\mu^{\nu}=0 \quad \forall \phi \in D(Q)
\end{equation}

Notice that $Q$ is symmetric when defined on $FC_b^\infty$:
\begin{equation}\label{Qsymm}
 \int Q\phi\ \psi \ d\mu^{\nu}=\int \phi\ Q \psi \ d\mu^{\nu}
\qquad \forall \phi, \psi \in FC^\infty_b,
\end{equation}
by direct computation, using integration by parts.
Therefore the
semigroup $R_t$ is symmetric;
hence we can define $R_t$ in any space 
$\mathcal L^p_{\mu^\nu}$ ($1\le p \le \infty$); first by the
$\mathcal L^1_{\mu^\nu}-\mathcal L^\infty_{\mu^\nu}$ duality
we define
it uniquely in $\mathcal L^1_{\mu^\nu}$ and then by
Riesz-Thorin theorem in $\mathcal L^p_{\mu^\nu}$ for $1< p < \infty$.
For simplicity, let us work in the Hilbert setting
of $\mathcal L^2_{\mu^\nu}$.

In particular, taking $\psi=1$ in \eqref{Qsymm} we get
\begin{equation}\label{ultimo}
 \int Q\phi\ d\mu^{\nu}=0 \qquad \forall \phi \in FC^\infty_b.
\end{equation}
This implies \eqref{Q-inv} if $FC^\infty_b$
is dense in $D(Q)$ with respect to the $Q$-operator
norm, or equivalently if $(Q, FC^\infty_b)$
is  essentially self-ajoint. The fact that $FC^\infty_b$
is a core for the infinitesimal generator of the semigroup
$R_t$ is proved by means of Theorem X.40 of \cite{rs}
since the semigroup is given by \eqref{semR} with \eqref{sol-z}.

As far as the nonlinear equation \eqref{eul} is concerned,
we have that the measure $\mu^\nu$ is infinitesimally invariant, that
is
\begin{equation}\label{invL}
  \int L\phi \ d\mu^{\nu}=0 \quad \forall \phi \in FC^\infty_b,
\end{equation}
where $(L, FC^\infty_b)$ is the Liouville operator associated to equation
\eqref{eul}, that is
\[
L\phi(x)=-\sum_{n,j}
 B_{n,j}(x,x) \frac{\partial \phi}{\partial x_{n,j}}(x).
\]
It is a linear operator in $\mathcal L_{\mu^\nu}^2$ with dense domain
$FC^\infty_b$ and is skew-symmetric, i.e.
\begin{equation}\label{skew}
  \int L\phi \ \psi \ d\mu^{\nu}= -\int \phi \ L\psi\ d\mu^{\nu}
        \quad \forall \phi,\psi \in FC^\infty_b.
\end{equation}
Indeed, integrating by parts
and noting that each $B_{n,j}$ does not depend on the variable $x_{n,j}$,
we have
\[\begin{split}
  \int L\phi  \ \psi \ d\mu^{\nu}&=
  - \int \sum_{n,j} B_{n,j}(x,x) \frac{\partial \phi}{\partial x_{n,j}}(x) \
  \psi(x) \ \mu^{\nu}(dx)
\\&=
  -\nu \int \underbrace{\sum_{n,j} B_{n,j}(x,x) x_{n,j}}_{\textstyle{=0 \text{ by
    } \eqref{Benerg}}} \ \phi(x)\ \psi(x) \ \mu^{\nu}(dx)
  \\&\;\;+ \int \phi(x) \ \sum_{n,j} B_{n,j}(x,x)
         \frac{\partial \psi}{\partial x_{n,j}}(x) \ \mu^{\nu}(dx)
  = - \int \phi \ L\psi   \ d\mu^{\nu}.
\end{split}\]
Taking $\psi=1$ in \eqref{skew}, we get \eqref{invL}.

Actually, if $\phi \in FC^1_b$ then $L\phi \in \mathcal L^q_{\mu^\nu}$
for any $q\ge 1$, since
the $B_{n,j}$'s are polynomials and $\mu^\nu$ is a Gaussian
measure.

As far as the nonlinear stochastic equation \eqref{equu} is
concerned,
its Kolmogorov operator is given by
\[\begin{split}
 K\phi(x)&=  \sum_n\sum_{j=1}^2\Big[
  k_n^2 \frac{\partial^2 \phi}{\partial x_{n,j}^2}(x)
    - B_{n,j}(x,x) \frac{\partial \phi}{\partial x_{n,j}}(x)
    - \nu k_n^2 x_{n,j}\frac {\partial \phi}{\partial x_{n,j}}(x)\Big]
\\&\equiv (Q+L)\phi
\end{split}\]
for $\phi \in FC^\infty_b$.
Again, given $\phi \in FC^2_b$ we have that
$ K\phi \in \mathcal L^q_{\mu^\mu}$  for any $q \ge 1$.
Hence we can consider $K$ as
a linear operator in $\mathcal L_{\mu^\nu}^q$ with dense domain
$FC^\infty_b$.

Since $K=Q+L$, from
\eqref{ultimo} and \eqref{invL} follows that
the measure $\mu^\nu$ is infinitesimally invariant for $K$, that
is
\begin{equation}\label{invK}
  \int K\phi \ d\mu^{\nu}=0 \quad \forall \phi \in FC^\infty_b.
\end{equation}

%SSSSSSSSSSSSSSSSSSSSSSSSSSSSSSSSSSS
\section{Stochastic viscous shell models}\label{S-viscoso}
Let us consider equation \eqref{equu}.
We are interested in solutions $u_x$ with initial data $x \in \mathbb H$.
To this end, we consider initial data in $H^{-\alpha}$ for
$0<\alpha<1$.
We assume in the whole section that
 the parameters $\nu>0$ and $\alpha \in ]0,1[$
  are fixed. The results hold true for arbitrary values of $\nu$ and
  $\alpha$ in the given range.

Here is our main result.
\begin{theorem}\label{TEOnu}
There exists a set $\tilde H\subset \mathbb H$ with $\mu^\nu(\tilde
H)=1$ such that for any $x \in \tilde H \cap H^{-\alpha} $
there exists a unique strong solution $u_x$ to equation \eqref{equu}  with
$u(0)=x$ and
\begin{equation}\label{potenzap}
 \mathbb E \| u_x\|_{C([0,T];H^{-\alpha})}^p<\infty
\end{equation}
for any $p\ge 1$ and any finite time interval $[0,T]$.
\\
Moreover, the measure $\mu^\nu$ is invariant for equation \eqref{equu}
in the sense that
\begin{equation}\label{u-x-inv}
 \int \mathbb E \phi(u_x(t))\ \mu^\nu(dx)= \int\phi(x)\ \mu^\nu(dx)
\end{equation}
for any $t\ge 0$ and $\phi\in \mathcal L^1_{\mu^\nu}(H^{-\alpha})$.
\end{theorem}

We prove this result in two steps, following
the lines of \cite{deb}. First, we construct a unique local
solution. Then, by means of an approximating problem we obtain an a
priori estimate, which provides the global existence.

\begin{remark}
The solution $u_x$ is a strong solution in the probabilistic sense and is
pathwise unique. It allows to construct the Markov  semigroup
$\{P_t\}_{t \ge 0}$ as
\[
 (P_t\phi)(x) = \mathbb E[\phi(u_x(t))].
\]
We have $P_t:B_b(H^{-\alpha})\to B_b(H^{-\alpha})$ for any $t\ge 0$,
where $B_b(H^{-\alpha})$ is the space of Borel bounded functions
$\phi: H^{-\alpha} \to \mathbb R$.
Actually,  $P_t\phi(x)$ is defined only for $\mu^\nu$-a.e. $x$, but this is
enough to give sense to \eqref{u-x-inv}.
In particular, thanks to \eqref{potenzap} $P_t$ is
 well defined on polynomial functions with values in
$\mathcal L^q_{\mu^\nu}$ for any $1\le q<\infty$; hence $P_t$
can be extended to $\mathcal L^q_{\mu^\nu}$.
\end{remark}

%SSSSSSSSSSSSSSSSSSSSSSSSSSSSSS
\subsection{Local existence}\label{S-loc-ex}
Consider  equation \eqref{equu} with initial data $u(0)=x \in H^{-\alpha}$.
We prove that it has a unique local solution, where the time interval on
which it is defined is random.
This basically relies on the fact that the equation has a locally Lipschitz
nonlinearity.
To this end, let us deal with the solution of equation
\eqref{equu} in the mild form (see, e.g., \cite{dpz})
\begin{equation}
 u_x(t)=e^{-\nu t A} x -\int_0^t e^{-\nu (t-s)A}B(u_x(s),u_x(s))\ ds +
       \int_0^t e^{-\nu (t-s)A}\sqrt{2A}\ dw(s).
\end{equation}

\begin{proposition}
Let $0<\alpha<1$.
For any $x \in H^{-\alpha} $ there exists a random time $\tau$ ($\tau
>0$ $\po$-a.s.) and
a unique process $u_x$  solving  equation \eqref{equu}  on the time
interval $[0,\tau]$ with initial value $x$ and such that
$$
 u_x \in C([0,\tau];H^{-\alpha}) \qquad \po-a.s.
$$
\end{proposition}
\proof
We use a fixed point theorem to prove that equation \eqref{equu} has
a local mild solution.
Set
\[
 z_0(t)=\int_0^t e^{-\nu (t-s)A}\sqrt{2A}\ dw(s).
\]
We know that a.a. the  paths of the process $z$ leave in
$C([0,T];\mathbb H)$.

We proceed pathwise. We define the mapping $\Psi$ as
\[
 (\Psi u)(t)= e^{-\nu t A} x -\int_0^t e^{-\nu (t-s)A}B(u(s),u(s))\ ds
 +z_0(t).
\]
We have that $\Psi:C([0,T];H^{-\alpha}) \to C([0,T];H^{-\alpha})$.
Indeed, $e^{-\nu t A} x$ and $z_0$ are in $C([0,T];H^{-\alpha})$;
we only need to  deal with the second term in the r.h.s.:
\[\label{stimeB1}
\begin{split}
 \sup_{0\le t \le T} \|\int_0^t & e^{-\nu (t-s)A} B(u(s),u(s))\ ds\|_{H^{-\alpha}}
\\ &\le
 \sup_{0\le t \le T} \int_0^t \|e^{-\nu (t-s)A}B(u(s),u(s))\|_{H^{-\alpha}} \ ds
\\
 &=
 \sup_{0\le t \le T} \int_0^t \|A^{\frac{1+\alpha}2}e^{-\nu (t-s)A}
                             A^{-\alpha-\frac 12}B(u(s),u(s))\|_{H^0} \ ds
\\
  &\le
  \sup_{0\le t \le T} \int_0^t
          \frac{c\|B(u(s),u(s))\|_{H^{-2\alpha-1}}}{(t-s)^{\frac{1+\alpha}2}}\ ds
  \qquad\text{ by } \eqref{semigr}
\\
  &\le c
  \sup_{0\le t \le T} \int_0^t
          \frac{\|u(s)\|^2_{H^{-\alpha}}}{(t-s)^{\frac{1+\alpha}2}}\ ds
   \qquad\text{ by } \eqref{stime}
\\
  &\le c
  \|u\|^2_{C([0,T];H^{-\alpha})}\sup_{0\le t \le T} \int_0^t
  \frac {ds}{(t-s)^{\frac{1+\alpha}2}}
\\
  & = C_0
  \|u\|^2_{C([0,T];H^{-\alpha})} T^{\frac {1-\alpha}2}
 \qquad \text{ for } \alpha<1.
\end{split}
\]
We have denoted by $C_0$ the latter constant, whereas in the previous
lines we used the same notation for different constants. From now on
we shall label only constants appearing in important relationships.

Hence, given  $x$ and $z_0$ we have
\begin{equation}\label{psi-lim}
 \|\Psi u\|_{L^\infty(0,T;H^{-\alpha})} \le
 \|x\|_{H^{-\alpha}}+\|z_0\|_{C([0,T];H^{-\alpha})}+ C_0
  \|u\|^2_{C([0,T];H^{-\alpha})} T^{\frac {1-\alpha}2}.
\end{equation}
The continuity in time is proved with similar estimates.

From \eqref{psi-lim} it follows that,
if $\|x\|_{H^{-\alpha}}+\|z_0\|_{C([0,T];H^{-\alpha})}< \frac R2$
and $T^{\frac{1-\alpha}2}< \frac 1{2C_0R}$,
then $\Psi$ maps the ball of radius $R$ of $C([0,T];H^{-\alpha})$
into itself:
\begin{align}
 \text{ for } &R>
 2\|x\|_{H^{-\alpha}}+2\|z_0\|_{C([0,T];H^{-\alpha})} \text{ and }\label{condiz}
\\  &\tau <
       (2C_0R)^{\frac 2{\alpha-1}},
   \label{condiz2}
\\    \text{ then}
 &\;\|\Psi u\|_{C([0,\tau];H^{-\alpha})} < R \quad\text{ if }
       \|u\|_{C([0,\tau];H^{-\alpha})} < R .
\end{align}
For instance, we can choose
\[
 R=3(\|x\|_{H^{-\alpha}}+\|z_0\|_{C([0,T];H^{-\alpha})}) \qquad
 \tau = [8 C_0 (\|x\|_{H^{-\alpha}}
  +\|z_0\|_{C([0,T];H^{-\alpha})})]^{\frac 2{\alpha-1}}.
\]
Since we proceed pathwise,  $R$ and $\tau$  are random variables,
almost surely positive and finite.

In addition we have that $\Psi$  is a contraction mapping
on the  ball of radius $R$ in $C([0,\tau];H^{-\alpha})$ for
$R$ and $\tau$ defined above, that is
\[
\begin{split}
 \forall x \in H^{-\alpha},\ & z_0 \in C([0,T];H^{-\alpha})\qquad
 \exists \ R>0 , \gamma <1, \tau >0:
\\
 & \|\Psi u^1-\Psi u^2\|_{C([0,\tau];H^{-\alpha})} < \gamma
 \|u^1-u^2\|_{C([0,\tau];H^{-\alpha})}
\\ &\qquad\qquad\qquad\qquad \forall
   \|u^1\|_{C([0,\tau];H^{-\alpha})}\le R, \|u^2\|_{C([0,\tau];H^{-\alpha})}\le R.
\end{split}
\]
To prove it, we use the bilinearity of the operator $B$
\[
 B(u^1,u^1)-B(u^2,u^2)=B(u^1, u^1-u^2) +B(u^1-u^2,u^2).
\]
As done before, we get
\[
\begin{split}
\|\Psi u^1&-\Psi u^2\|_{C([0,\tau];H^{-\alpha})}
  \le
\\&
  \sup_{0\le t \le \tau} \int_0^t
  \| e^{-\nu (t-s)A}[B(u^1(s), u^1(s)-u^2(s))
         +B(u^1(s)-u^2(s),u^2(s) )]\|_{H^{-\alpha}}\ ds
\\
  & \le
  C_0[\|u^1\|_{C([0,\tau];H^{-\alpha})}+\|u^2\|_{C([0,\tau];H^{-\alpha})}]
  \|u^1-u^2\|_{C([0,\tau];H^{-\alpha})} \tau^{\frac {1-\alpha}2}.
\end{split}
\]
The same choice of $R$ and $\tau$ as in \eqref{condiz}-\eqref{condiz2}
provides the result with $\gamma=2C_0R\tau^{\frac {1-\alpha}2}$.
\hfill $\Box$

\medskip
Given $x$ and $z_0$ we have obtained the solution $u_x$ on the time
interval $[0,\tau]$.
We could proceed beyond time 
$\tau$; we construct the solution $u_x$ starting at time
$\tau$ from $u_x(\tau)$. Our construction shows that if
$\|u(0)\|_{H^{-\alpha}}+ \|z_0\|_{}\le \frac R2$
then $\|u(\tau)\|_{H^{-\alpha}}\le R$; starting from time $\tau$ the
amplitude of the next time interval is smaller than $\tau$ and as
usual for nonlinear equations we are not granted to cover the whole
time interval $[0,T]$ by a repeated procedure.

To get a global solution we need an a priori estimate. This will be the
argument of the next two sections; we need to approximate the
nonlinear term $B$ and to use the invariance of the measure $\mu^\nu$
for the approximate problem. Then we recover the result for
equation \eqref{equu}.

%SSSSSSSSSSSSSSSSSSSSSSSSSSSSSSSSSSSSSSSSSss
\subsection{Finite dimensional approximation of $B$}
For any $M \in \mathbb N$,
let $\Pi_M$ be the projection operator in $H^0$ defined as
$\Pi_M(u_1,u_2, \ldots)=(u_1,u_2, \ldots,u_M, 0, 0, \ldots)$.

Moreover, for $M\ge 3$
let $B^M$ be the bilinear operator defined
as
\[
 B^M(u,v)=\Pi_M B(\Pi_M u, \Pi_M v).
\]
$B^M$ is a bounded operator
from $(\mathbb R^2)^M \times (\mathbb R^2)^M $ to $(\mathbb R^2)^M
$. In addition we have the same result of Lemma \ref{Bgenerale}
\begin{equation}\label{BMspazi}
 \|B^M(u,v)\|_{H^{-\alpha_3}}\le c \|u\|_{H^{\alpha_1}} \|v\|_{H^{\alpha_2}}\qquad
 \text{ if }  \alpha_1+\alpha_2+\alpha_3=1,
\end{equation}
where the constant $c$ is independent of $M$.

We have the relationship corresponding to \eqref{Benerg}:
\begin{equation}\label{energia}
 \sum_{n=1}^M B^M_n(u,v)\cdot v_n=0
\end{equation}

The  approximation problem associated to \eqref{equu} is
\begin{equation}\label{gal}
\begin{cases}
du^M(t)+[\nu Au^M(t)+B^M(u^M(t),u^M(t))]dt = \Pi_M \sqrt {2A}\ dw(t)\\
 u^M(0)=\Pi_M x
\end{cases}
\end{equation}
We consider any finite time interval $[0,T]$ and set
$\mu^{\nu,M}=\otimes_{n=1}^M \mu^\nu_n$.

In order to study this problem, we make precise some properties of the
bilinear term $B^M$.
\begin{lemma}
\begin{equation}\label{stimeM}
 \sup_M \int_{\mathbb H} \|B^M(x,x)\|^p_{H^{-1-\beta}} \mu^\nu(dx)
\le \int_{\mathbb H} \|B(x,x)\|^p_{H^{-1-\beta}}
<\infty
\end{equation}
for any $\beta>0$.

Moreover, for any $p\ge 1$
\begin{equation}\label{binf}
\lim_{M \to \infty} \|B^M(u,u)-B(u,u)\|_{L^p(0,T;H^{-1-2\alpha})} =0
       \qquad\text{ if } u \in C([0,T];H^{-\alpha}).
\end{equation}
\end{lemma}
\proof
\eqref{stimeM} is proved as in Proposition \ref{Bstaz}.

We give details for \eqref{binf} in the case of the SABRA model.
First,
\[
 B^M_{n,1}(x,x)-B_{n,1}(x,x)=
 \begin{cases}
 0 & \text{ for } n \le M-2\\
 -a k_M (x_{M,1}x_{M+1,2} - x_{M,2}x_{M+1,1}) & \text{ for } n=M-1\\
 -a k_{M+1}(x_{M+1,1}x_{M+2,2} - x_{M+1,2}x_{M+2,1}) &\\[-2mm]
      & \text{ for } n=M\\[-2mm]
     \;\qquad  -b k_M (x_{M-1,1}x_{M+1,2} - x_{M-1,2}x_{M+1,1}) & \\
 -B_{n,1}(x,x) & \text{ for } n \ge M+1
 \end{cases}
\]
and
\[
 B^M_{n,2}(x,x)-B_{n,2}(x,x)=
 \begin{cases}
 0 & \text{ for } n \le M-2\\
 -a k_M (-x_{M,1}x_{M+1,1} - x_{M,2}x_{M+1,2})  & \text{ for } n=M-1\\
 -a k_{M+1}(-x_{M+1,1}x_{M+2,1} - x_{M+1,2}x_{M+2,2})&\\[-2mm]
      & \text{ for } n=M\\[-2mm]
  \;\qquad -b k_M (-x_{M-1,1}x_{M+1,1} - x_{M-1,2}x_{M+1,2}) &\\
 -B_{n,2}(x,x) & \text{ for } n \ge M+1
 \end{cases}
\]
Therefore
\[\begin{split}
 \|B^M(x,x)-B(x,x)\|_{H^{-1-2\alpha}}
 &\le
 |B_{M-1}^M(x,x)-B_{M-1}(x,x)|^2 k_{M-1}^{2(-1-2\alpha)}
 \\&+ |B_{M}^M(x,x)-B_{M}(x,x)|^2 k_M^{2(-1-2\alpha)}
 \\&+ \sum_{n=M+1}^\infty |B_n(x,x)|^2 k_n^{2(-1-2\alpha)}.
 \end{split}
\]
Following the proof  of Lemma \ref{Bgenerale}
we get that for any $x \in H^{-\alpha}$ we have
\[
 \|B^M(x,x)-B(x,x)\|_{H^{-1-2\alpha}} \le
 c \|(\Pi_{M-2}-I)x\|^2_{H^{-\alpha}}
 \to 0
       \qquad \text{ as } M \to \infty,
\]
Hence,
\[
 \int_0^T \|B^M(u(s),u(s))-B(u(s),u(s))\|^p_{H^{-\alpha}} ds
 \le c \int_0^T \|\Pi_{M-2}u(s)-u(s)\|^{2p}_{H^{-\alpha}}ds.
\]
Moreover,
\[
  \lim_{N \to \infty} \|\Pi_{N}u(s)-u(s)\|^{2p}_{H^{-\alpha}}=0
\]
for every $s$ if $u \in C([0,T];H^{-\alpha})$, and
\[
  \int_0^T \|\Pi_{N}u(s)-u(s)\|^{2p}_{H^{-\alpha}}ds \le
  \int_0^T \|u(s)\|^{2p}_{H^{-\alpha}}ds
 \qquad \forall N.
\]
We conclude by dominated convergence.\hfill $\Box$

\medskip
For equation \eqref{gal} we have the following standard  result.
\begin{proposition}\label{pro-gal}
For each $x \in (\mathbb R^2)^\infty$ and $M$,
there exists a unique strong
solution to equation \eqref{gal}, which is a continuous and Markov process.
\\
Moreover, $\mu^{\nu,M}$ is the unique invariant measure for equation \eqref{gal}.
\\
Finally, there exists a unique stationary solution, which is a
$\mu^{\nu,M}$-stationary process.
\end{proposition}
\proof
The existence and uniqueness result  is  standard; indeed, equation \eqref{gal}
is an evolution equation in the state space $({\mathbb R}^2)^M$.
In the
finite dimensional case, the equation with a locally Lipschitz nonlinearity has
a unique local solution, defined on a random time interval
$[0,\tau]\subseteq [0,T]$ (see, e.g., \cite{protter}).
The global solution, that is the solution existing on the
whole time interval $[0,T]$, is shown to exist thanks to an a priori
estimate obtained by It\^o formula for $d\|u^M(t)\|^2_{H^0}$,
using property \eqref{energia}:
\[
 \mathbb E \sup_{0\le t \le T} \|u^M(t)\|^2_{H^0}\le \|\Pi_M x\|_{H^0}^2+
 t \ C(M)
\]
where $C(M)$ is a constant quantity depending on $M$
(see, e.g., \cite{ac}  for the similar case of the 2D Navier-Stokes equation).

Moreover, for the finite dimensional equation \eqref{gal} existence
and uniqueness of an invariant
measure hold true, since the noise is non-degenerate i.e. it acts on all the
components (see, e.g.,  \cite{khas}).
Let us prove that this unique invariant measure is indeed
$\mu^{\nu,M}$.
Denote by $u_x^M(t)$ the unique solution of \eqref{gal} started at time 0
from $x\in ({\mathbb R}^2)^M$ and evaluated at time $t$.
This uniquely defines a Markov semigroup $\{P^M_t\}_{t\ge 0}$:
\[
 (P^M_t \phi) (x)=\mathbb E[\phi(u_x^M(t))], \;
 \phi \in B_b(({\mathbb R}^2)^M).
\]
Actually, the
semigroup can be defined as acting in $\mathcal L^1_{\mu^{\nu,M}}$,
as we shall see in the following lines.

We now prove that the measure
$\mu^{\nu,M}$ is an invariant measure for
\eqref{gal} in the sense that
\begin{equation}\label{P-inv}
 \int P_t^M \phi\ d\mu^{\nu,M}= \int \phi\ d\mu^{\nu,M}
  \qquad \forall  t\ge 0, \ \phi\in \mathcal L^1_{\mu^{\nu,M}}.
\end{equation}

The invariance \eqref{P-inv} is equivalent to the infinitesimal invariance
\begin{equation}\label{KM-inv}
 \int K^M\phi\ d\mu^{\nu,M}=0 \qquad
  \forall \phi \in D(K^M)\subset \mathcal L^1_{\mu^{\nu,M}}
\end{equation}
being $K^M:D(K^M)\to \mathcal L^1_{\mu^{\nu,M}}$
the infinitesimal generator of the  semigroup $P^M_t$ in
$\mathcal L^1_{\mu^{\nu,M}}$.

On the set $C^\infty_b(({\mathbb R}^2)^M )$) of infinitely
differentiable functions bounded with all derivatives bounded,
the operator $K^M$ has the expression
\begin{equation}
K^M\phi(x)=  \sum_{n=1}^M\sum_{j=1}^2\Big[
  k_n^2 \frac{\partial^2 \phi}{\partial x_{n,j}^2}(x)
    - B^M_{n,j}(x,x) \frac{\partial \phi}{\partial x_{n,j}}(x)
    - \nu k_n^2 x_{n,j}\frac {\partial \phi}{\partial x_{n,j}}(x)\Big].
\end{equation}
First, we have that
\begin{equation}\label{ultimo2}
 \int K^M\phi\ d\mu^{\nu,M}=0 \qquad \forall \phi \in
C^\infty_b(({\mathbb R}^2)^M ).
\end{equation}
Indeed, if $\phi \in C^\infty_b(({\mathbb R}^2)^M )$
then $K^M\phi$ is $\mu^{\nu,M}$-integrable; morevoer,
\[\begin{split}
 \int K^M\phi\ & d\mu^{\nu,M}
\\&=
 \int  \sum_{n,j} \Big[k_n^2 \frac{\partial^2 \phi}{\partial x_{n,j}^2}(x)
-B^M_{n,j}(x,x) \frac{\partial \phi}{\partial x_{n,j}}(x)
    - \nu k_n^2 x_{n,j}\frac {\partial \phi}{\partial
      x_{n,j}}(x)\Big]\ \mu^{\nu,M}(dx)
\intertext{and integrating by parts}
\\&=
  - \nu \int  \sum_{n,j}B^M_{n,j}(x,x) x_{n,j} \phi(x)\ \mu^{\nu,M}(dx)
 =0 \;\;\text{ by } \eqref{energia}.
\end{split}
\]

Secondly,
\[
K^M\phi(x)\equiv Q^M\phi(x)- \sum_{n=1}^M\sum_{j=1}^2
  B^M_{n,j}(x,x) \frac{\partial \phi}{\partial x_{n,j}}(x)
\]
As done in Section \ref{S-inva} for the Ornstein--Uhlenbeck operator $Q$,
we can prove that $ C^\infty_b(({\mathbb R}^2)^M )$
is a domain of essential self-adjointness for $Q^M$.
Using  Theorem 1.2 of \cite{s} we get that
$C^\infty_b(({\mathbb R}^2)^M )$ is a core for $K^M$, that is
$C^\infty_b(({\mathbb R}^2)^M)$
is dense in $D(K^M)$ with respect to the graph norm.

From this density result  we have that \eqref{ultimo2} implies
\eqref{KM-inv}.
\hfill$\Box$

\medskip
Now, define
\begin{equation}\label{solvM}
 v_x^M=u^M_x+(I-\Pi_M)z_x.
\end{equation}
This process is  a solution of
\begin{equation}\label{inf-gal}
\begin{cases}
 dv_x^M(t) +[\nu Av_x^M(t)+B^M(v_x^M(t),v_x^M(t))]dt = \sqrt {2A} \ dw(t)\\
 v_x^M(0)=x
\end{cases}
\end{equation}

We have
\begin{proposition}
For each $x \in H^{-\alpha}$ and $M$,
there exists a unique strong
solution $v^M_x$ to equation \eqref{inf-gal},
which is a Markov process and for any $T>0$ and finite
\[
  v^M_x \in C([0,T];H^{-\alpha}) \qquad \po-a.s.
\]
for any finite time interval $[0,T]$. Moreover
\begin{equation}\label{Muniformi}
 \sup_M \int \mathbb E\|v^M_x\|^p_{C([0,T];H^{-\alpha})}\mu^\nu(dx) <\infty
\end{equation}
for any $p\ge 1$.
\\
The measure $\mu^{\nu}$ is the unique invariant measure for equation
\eqref{gal}, that is
\begin{equation}\label{Minvariance}
 \int \mathbb E \phi (v^M_x(t))\ \mu^\nu(dx)= \int \phi(x) \ \mu^\nu(dx)
\end{equation}
for any $t \ge 0$ and $\phi \in \mathcal L^1_{\mu^\nu}$.
\\
\end{proposition}
\proof
According to Proposition \ref{pro-gal}
and to the results of Section \ref{S-equ},
we have that for any $x \in H^{-\alpha}$ there exists a unique
solution of \eqref{inf-gal}
with paths in $C([0,T];H^{-\alpha})$. This is therefore given by \eqref{solvM}.
Again we prove the uniqueness of the invariant measure dealing
separately with the dynamics on the first $M$ modes and on the
remaining modes.

The invariance \eqref{Minvariance}
is proved as in the proof of Proposition \ref{pro-gal}, since the
Kolmogorov operator associated to equation \eqref{inf-gal} is
\[
  Q- \sum_{n=1}^M\sum_{j=1}^2
  B^M_{n,j} \frac{\partial \;\;}{\partial x_{n,j}}.
\]

What remains to be proved is \eqref{Muniformi}.
Consider the mild form of the solution to equation \eqref{inf-gal}
\[
 v^M_x(t)=z_x(t)-\int_0^t e^{-\nu (t-s)A}B^M(v^M_x(s),v^M_x(s))\ ds
\]
Then
\begin{multline}\label{stimeconp}
 \|v^M_x\|^p_{C([0,T];H^{-\alpha})}
\le 2^{p-1}
   \|z_x\|^p_{C([0,T];H^{-\alpha})}
\\+ 2^{p-1}
   (\sup_{0\le t \le T}
      \|\int_0^t e^{-\nu (t-s)A}B^M(v^M_x(s),v^M_x(s))\ ds\|_{H^{-\alpha}})^p.
\end{multline}
We estimate the latter term;
\[
\begin{split}
\|&\int_0^t e^{-\nu (t-s)A}B^M(v^M_x(s),v^M_x(s))\ ds\|_{H^{-\alpha}}
\\
&\le
  \sup_{0\le t \le T}
      \int_0^t \|e^{-\nu
        (t-s)A}B^M(v^M_x(s),v^M_x(s))\|_{H^{-\alpha}}\ ds
\\&=
 \sup_{0\le t \le T}
      \int_0^t \|A^{\frac 12}e^{-\nu
        (t-s)A} A^{-\frac{1+\alpha}2}B^M(v^M_x(s),v^M_x(s))\|_{H^0}\ ds
\\&\le c
 \sup_{0\le t \le T}
     \int_0^t \frac1{(t-s)^{\frac 12}}\|B^M(v^M_x(s),v^M_x(s))\|_{H^{-1-\alpha}}\ ds
  \qquad\text{ by } \eqref{semigr}
\\&\le c
  \sup_{0\le t \le T}
     \left(\int_0^t \frac{ds}{(t-s)^{\frac 34}} \right)^{\frac 23}
     \left(\int_0^t
     \|B^M(v^M_x(s),v^M_x(s))\|^3_{H^{-1-\alpha}}\ ds\right)^{\frac 13}
 \text{ by H\"older ineq.}
\\&=
  c (4T^{1/4})^{\frac 23}
  \left(\int_0^T
  \|B^M(v^M_x(s),v^M_x(s))\|^3_{H^{-1-\alpha}}\ ds\right)^{\frac 13} .
\end{split}
\]

Now in \eqref{stimeconp}
 we take the integral with respect to $\po$ and $\mu^\nu$.
Using the invariance of the measure $\mu^\nu$, we get
\[\begin{split}
 \int \mathbb E  &\|v^M_x\|^p_{C([0,T];H^{-\alpha})} \ \mu^\nu(dx)
\\&\le
 c \int \mathbb E  \|z_x\|^p_{C([0,T];H^{-\alpha})} \ \mu^\nu(dx)
+c T^{\frac p6}\left(T \int \|B^M(x,x)\|^{3p}_{H^{-1-\alpha}}
\ \mu^\nu(dx)\right)^{\frac p3}.
\end{split}\]
Finally,  the uniform estimate \eqref{Muniformi}
follows from the uniform estimate \eqref{stimeM}.
\hfill$\Box$

\begin{remark}
From the properties of equations \eqref{gal} and \eqref{equz},
we also have that
there  exists a unique stationary solution
of \eqref{inf-gal}, which is a $\mu^\nu$-stationary process.
\end{remark}

%SSSSSSSSSSSSSSSSSSSSSSSSSSSSSS
\subsection{Global existence}
Let us come back to the local existence result.
Consider the solutions $v^M_x$ as living in $C([0,\tau];H^{-\alpha})$.
Estimates similar to those of  Section
\ref{S-loc-ex} allow to prove that
the whole sequence $v_x^M$ converges (pathwise) to $u_x$ in the
$C([0,\tau];H^{-\alpha})$-norm.

\begin{proposition}
\begin{equation}\label{gal-conv}
  \lim _{M \to \infty}\|u_x-v^M_x\|_{C([0,\tau];H^{-\alpha})} =0 \qquad\po-a.s.
\end{equation}
\end{proposition}
\proof
We have
\[
 u_x(t)-v^M_x(t)=
\int_0^t e^{-\nu (t-s)A} [B^M(v^M_x(s),v^M_x(s))-B(u_x(s),u_x(s))]ds.
\]
Moreover
\[
 B^M(v^M,v^M)-B(u,u)=B^M(v^M,v^M-u)+B^M(v^M-u,u)+B^M(u,u)-B(u,u);
\]
therefore
\begin{multline*}
\|u_x-v^M_x\|_{C([0,\tau];H^{-\alpha})}\\
 \le
 \sup_{0\le t\le \tau} \int_0^t  \|e^{-\nu (t-s)A}
  [B^M(v^M_x(s),v^M_x(s)-u_x(s))+B^M(v^M_x(s)-u_x(s),u_x(s))]\|_{H^{-\alpha}}ds
\\
+ \sup_{0\le t\le \tau} \int_0^t
  \|e^{-\nu (t-s)A}[B^M(u_x(s),u_x(s))-B(u_x(s),u_x(s))]\|_{H^{-\alpha}}ds.
\end{multline*}
We estimate
\[
 \|B^M(v^M,v^M-u)\|_{H^{-1-2\alpha}}+\|B^M(v^M-u,u)\|_{H^{-1-2\alpha}}
\le c (\|v^M\|_{H^{-\alpha}}+\|u\|_{H^{-\alpha}})   \|v^M-u\|_{H^{-\alpha}}
\]
as done in \eqref{stime}. Moreover
\[
\begin{split}
 \int_0^t &\|e^{-\nu (t-s)A}[B^M(u_x(s),u_x(s))-B(u_x(s),u_x(s))]\|_{H^{-\alpha}}ds
\\&\le
 \int_0^t \frac c{(t-s)^{\frac{1+\alpha}2}}
     \|B^M(u_x(s),u_x(s))-B(u_x(s),u_x(s))\|_{H^{-1-2\alpha}}ds \quad \text{
       by } \eqref{semigr}
\\&\le
 \left(c\int_0^t\frac{ds} {(t-s)^{\frac{1+\alpha}{1+\sqrt\alpha}}}\right)
         ^{\frac{1+\sqrt\alpha}2}
 \left(\int_0^t \|B^M(u_x(s),u_x(s))-B(u_x(s),u_x(s))\|
    ^{\frac2{1-\sqrt\alpha}}_{H^{-1-2\alpha}}ds\right)^{\frac {1-\sqrt\alpha}2}
\end{split}
\]
by H\"older inequality.

With usual computations we get
\[
\begin{split}
 \| u_x-v^M_x\|_{C([0,\tau];H^{-\alpha})}
&\le
 C_0 \tau^{\frac{1-\alpha}2}
 (\|v_x^M\|_{C([0,\tau];H^{-\alpha})}+\|u_x\|_{C([0,\tau];H^{-\alpha})})
  \|v^M_x-u_x\|_{C([0,\tau];H^{-\alpha})}
\\&\;\;+
  C_1 \tau^{\frac{\sqrt \alpha-\alpha}2} \|B^M(u_x,u_x)-B(u_x,u_x)\|
    _{L^{\frac2{1-\sqrt\alpha}}(0,\tau;H^{-1-2\alpha})}
\\&\le
  2C_0 R \tau^{\frac{1-\alpha}2} \|v^M-u\|_{C([0,\tau];H^{-\alpha})}
\\&\;\;+
  C_1 \tau^{\frac{\sqrt \alpha-\alpha}2} \|B^M(u_x,u_x)-B(u_x,u_x)\|
    _{L^{\frac2{1-\sqrt\alpha}}(0,\tau;H^{-1-2\alpha})}
\end{split}\]
choosing $R$ and $\tau$ as in \eqref{condiz}-\eqref{condiz2}.
Therefore
\[
 \underbrace{[1-2C_0 R \tau^{\frac{1-\alpha}2}]}_{>0}
 \|u_x-v^M_x\|_{C([0,\tau];H^{-\alpha})}
 \le
  C_1   \tau^{\frac{\sqrt \alpha-\alpha}2}
 \underbrace{\|B^M(u_x,u_x)-B(u_x,u_x)\|_{L^{\frac2{1-\sqrt\alpha}}(0,\tau;H^{-1-2\alpha})}
    }_{\textstyle{
  \to 0 \text{ by } \eqref{binf}}}.
\]
This implies \eqref{gal-conv}.
\hfill $\Box$

\medskip
On the other hand,
\eqref{Muniformi}
says that the sequence $\{v^M_x\}_M$ is bounded in
$L^p(\Omega\times H^{-\alpha}, \po\otimes \mu^\nu;C([0,T];H^{-\alpha})) $.
Therefore, by the Banach-Alaoglu theorem
there exists a subsequence  $\{v^{M_i}_x\}_i$ $\star$-weakly converging
in $L^p(\Omega\times H^{-\alpha}, \po\otimes \mu^\nu;C([0,T];H^{-\alpha})) $
 to some $v_x$; moreover, the limit satisfies
\[
 \int \mathbb E \|v_x\|^p_{C([0,T];H^{-\alpha})}\ \mu^\nu(dx)<\infty.
\]
In particular, for $\mu^\nu$-a.e. $x$
\begin{equation}
  \|v_x\|^p_{C([0,T];H^{-\alpha})}<\infty \qquad \po-a.s.
\end{equation}
Since we also know that the whole sequence  $\{v^M_x\}_M$
converges pathwise to $u_x$ in $C([0,\tau];H^{-\alpha}) $ we have
\[
  u_x|_{[0,\tau]}=v_x|_{[0,\tau]} \qquad \po-a.s.
\]
and
\[
 \|u_x\|_{C([0,\tau];H^{-\alpha})} \le \|v_x\|_{C([0,T];H^{-\alpha})} \qquad \po-a.s.
\]
Now, by means of this bound
we get that the path $u_x$ exists on  the time interval
$[0,T]$.
Indeed,
if we choose $\tau< (2C_0R)^{\frac 2{\alpha-1}}$
with $R>
 2\|v_x\|_{C([0,T];H^{-\alpha})}+2\|z_0\|_{C([0,T];H^{-\alpha})}$,
we can repeat the construction of the path
$u_x$ on the time interval $[\tau, 2\tau]$ and so on until we recover
the whole time interval $[0,T]$.
This gives \eqref{potenzap}.

As far as the invariance of the measure $\mu^\nu$ is concerned,
notice that we proved  that
\[
   \lim _{M \to \infty}\|u_x-v^M_x\|_{C([0,T];H^{-\alpha})} =0 \qquad\po-a.s.
\]
Hence, given $\phi \in C(H^{-\alpha})$
\[
  \lim _{M \to \infty}\phi(v^M_x(t))=\phi(u_x(t)) \qquad\po-a.s.
\]
for any $t$. From \eqref{Minvariance}
and by the dominated convergence
theorem we get that for any $\phi \in C_b(H^{-\alpha})$
\[
 \int \mathbb E \phi (u_x(t))\ \mu^\nu(dx)= \int \phi(x)
 \ \mu^\nu(dx), \qquad \forall t \ge 0
\]
that is
\[
 \int P_t \phi \ d\mu^\nu= \int \phi  \ d\mu^\nu, \qquad \forall t \ge 0.
\]

Because of \eqref{potenzap}
we can extend this property to any $\phi \in \mathcal L^q_{\mu^\nu}$
($1\le q < \infty$) and we get the invariance \eqref{u-x-inv}.

\begin{remark}
We point out that we have not proved that $FC^\infty_b$
is a core for the infinitesimal generator of the semigroup
$P_t$ in $\mathcal L^1_{\mu^\nu}$, whereas we proved before (for
any $M$) that $FC^\infty_b$
is a core for the infinitesimal generator of the semigroup
$P^M_t$ in $\mathcal L^1_{\mu^\nu}$.
The criterium of \cite{s} used in the
previous sections (with the quadratic term $B^M$)
does not work for the operator $K$ (with the ''full'' quadratic term
$B$).
There are few methods to prove this kind of results for
infinite-dimensional non-symmetric operators; also the
approximative approach of Eberle (see \cite{e} Chapter 5) is not helpful.
However,thanks to the pathwise uniqueness  we have proved
the strong Markov uniqueness of the Kolmogorov operator
$(K,FC^\infty_b)$  in $\mathcal L^1_{\mu^\nu}$ (with the terminology of \cite{e}).
Actually, the result of strong Markov uniqueness
 would be true with respect to any invariant
measure for equation \eqref{equu} whose support is included in
$H^{-\alpha}$ for some $0<\alpha<1$.
\end{remark}

We conclude the analysis of the existence and uniqueness result, by noting
that when we consider as initial data a random variable independent
of $w$ and with law $\mu^\nu$,
we can construct a unique strong solution $u_{st}$ with the same technique of
Theorem \ref{TEOnu}; indeed the initial data $u_{st}(0)$ is such that
$\|u_{st}(0)\|_{\mathbb H}^2<\infty$ a.s..
This solution is the limit of the $\mu^\nu$-stationary Galerkin
approximations $v^M_{st}$
\[
 \begin{cases}
 dv_{st}^M(t) +[\nu Av_{st}^M(t)+B^M(v_{st}^M(t),v_{st}^M(t))]dt
             = \sqrt {2A} \ dw(t),\qquad t>0\\
 v_{st}^M(0)\text{ has law } \mu^\nu
\end{cases}
\]
Therefore  we get that also this solution is a
$\mu^\nu$-stationary process.
The only difference in the proof is that
\eqref{Muniformi} is replaced by
\[
 \sup_M  \mathbb E\|v^M_{st}\|^p_{C([0,T];H^{-\alpha})} <\infty
\]
and \eqref{Minvariance}  by
\[
  \mathbb E \phi (v^M_{st}(t)) = \int \phi(x) \ \mu^\nu(dx)
  \qquad \forall t\ge 0.
\]
Hence we have
\begin{proposition}\label{pro-staz}
If $u(0)$ is a random variable with law $\mu^\nu$ and independent of
$w$, there exists a unique strong solution to equation
\eqref{equu} with paths in
$C([0,\infty);\mathbb H)$ $\po$-a.s.. This is a $\mu^\nu$-stationary process.
\end{proposition}

This allows to get \eqref{u-x-inv} directly.

%SSSSSSSSSSSSSSSSSSSSSSSSSSSSSSSSSSS
\section{Inviscid shell models}\label{S-eul}
Consider equation \eqref{eul}.
It has been studied considering initial data of finite energy or even
more regular (see \cite{clt2,bbbf,bm} and the
references therein).
However, we are interested in solutions
having $\mu^\nu$ as invariant measure. This requires to deal with
initial data in the space $\mathbb H$ but not in the space $H^0$.

Notice that equation \eqref{eul}  is obtained
from the viscous stochastic shell model \eqref{equu}
by neglecting the viscous and the stochastic terms.
For this reason, for any $\ep>0$ let us consider the equation
\begin{equation}\label{eq-ep}
 du^\ep(t)+[\nu \ep A u^\ep(t) + B(u^\ep(t),u^\ep(t))]dt
=  \sqrt {2\ep A} \ dw(t), \qquad t>0.
\end{equation}
For $\ep=0$ this reduces to equation \eqref{eul}.

To analyse equation \eqref{eq-ep}
we can apply the results of the previous section; they hold true for
any $\ep>0$. The fact that the measure $\mu^\nu$ is an invariant
measure for any $\ep>0$ can be easily checked by looking at the
expression of the Kolmogorov operator associated to equation \eqref{eq-ep}:
$K^\ep=\ep Q+L$.
Therefore, according to Proposition \ref{pro-staz}
equation \eqref{eq-ep}
has a unique $\mu^\nu$-stationary solution $\overline v^{\nu,\ep}$;
this process is a strong solution
and has paths in $C([0,\infty);H^{-\alpha})$ ($\alpha>0$) a.s..

We are going to prove that there exists a subsequence
$\{\overline v^{\nu,\ep_n}\}_n$ converging in a
suitable sense as $\ep_n\to 0$
to a process which solves \eqref{eul} for $t\ge 0$.
First, we have
\begin{proposition}
For any $0\le \tilde \beta <\frac 12$, $\tilde \alpha>0$ and $T>0$,
the family $\{\overline v^{\nu,\ep}\}_{0<\ep\le 1}$
is tight in $C^{\tilde\beta}([0,T];H^{-2-\tilde\alpha})$.
\end{proposition}
\proof
We
write equation \eqref{eq-ep} in the integral form:
\begin{equation}\label{leqint}
 \overline v^{\nu,\ep}(t)=\overline v^{\nu,\ep}(0)
 - \nu \ep \int_0^t A \overline v^{\nu,\ep}(s)\ ds
 - \int_0^t B(\overline v^{\nu,\ep}(s),\overline v^{\nu,\ep}(s))\ ds
 + \sqrt {2\ep A} w(t).
\end{equation}
$\overline v^{\nu,\ep}(t)$ and $\overline v^{\nu,\ep}(0)$ are
random variables with  law
$\mu^\nu$. We estimate the latter three terms.

First, for $\mu^\nu$-stationary processes we have
\begin{equation}\label{nu-A}
\begin{split}
 \mathbb E\Big\|\int_0^\cdot A\overline v^{\nu,\ep}(s) &
   ds\Big\|^p_{W^{1,p}(0,T;H^{-2-\alpha})}
\\& \le (1+T^p)\int_0^T \mathbb E[\|\overline v^{\nu,\ep}(s)\|^p_{H^{-\alpha}} ] ds
\\
& = T(1+T^p) \int_{\mathbb H} \|x\|^p_{H^{-\alpha}} \mu^\nu(dx) =:
\tilde C_p
\end{split}\end{equation}
\begin{equation}\label{nu-B}
\begin{split}
 \mathbb E \Big\|\int_0^\cdot &
   B  (\overline v^{\nu,\ep}(s),\overline v^{\nu,\ep}(s)) ds
     \Big\|^p_{W^{1,p}(0,T;H^{-1-\alpha})}
\\
& \le  (1+T^p)\int_0^T \mathbb E
      [\|B(\overline v^{\nu,\ep}(s),\overline v^{\nu,\ep}(s))\|^p_{H^{-1-\alpha}} ] ds
\\
& = T(1+T^p) \int_{\mathbb H} \|B(x,x)\|^p_{H^{-1-\alpha}} \mu^\nu(dx)
\end{split}\end{equation}

Moreover, for any $0\le \beta < \frac 12$, $\alpha>0$
\begin{equation}\label{nu-W}
 \mathbb E[\|\sqrt A w\|_{C^\beta([0,T];H^{-1-\alpha})}]
 \le  \overline C_\beta \quad \text{ by } \eqref{holder}.
\end{equation}

Since $\ep\le 1$,
we get
\[
 \sup_{0 < \ep \le 1}\mathbb E\Big\|\int_0^\cdot \nu \ep A\overline v^{\nu,\ep}(s)
   ds\Big\|^p_{W^{1,p}(0,T;H^{-2-\alpha})}
 \le \nu^p \tilde C_p
\]
and
\[
\sup_{0 < \ep \le 1}
 \mathbb E[\|\sqrt {2\ep A} w\|_{C^\beta([0,T];H^{-1-\alpha})}]
 \le  \sqrt {2}  \overline C_\beta
\]

Now, we use that $W^{1,p}(0,T)\subset C^\beta([0,T])$ if  $1-\frac 1p>\beta$.
Then, using the previous estimates in  \eqref{leqint},
given any $0\le\beta<\frac 12$,
$p>\frac 1{1-\beta}$  and $\alpha>0$ we have 
\begin{equation}\label{tigh}
  \sup_{0 < \ep \le 1}
  \mathbb E[\|\overline v^{\nu,\ep}\|^p_{C^\beta([0,T];H^{-2-\alpha})}]<\infty .
\end{equation}
On the other hand, the space $C^\beta([0,T];H^{-1-\alpha})$ is
compactly embedded in $C^{\tilde \beta}([0,T];H^{-1-\tilde \alpha})$
if $\tilde \alpha >\alpha, \tilde \beta <\beta$; this follows from the
compact embedding
$H^{-1-\alpha}\Subset H^{-1-\tilde \alpha}$
and from the Ascoli-Arzel\`a theorem.

The tightness  follows from \eqref{tigh} as usual  by means of
Chebyshev inequality, since $\alpha$ and
$\beta$ are arbitrary values with the stated restrictions.
\hfill$\Box$

\medskip
Similarly we work on the time interval $[-T,0]$ by considering
the reversed-time parabolic nonlinear equation
\begin{equation}\label{eq--ep}
 du^\ep(t)+[-\nu \ep A u^\ep(t) + B(u^\ep(t),u^\ep(t))]dt
=  \sqrt {2\ep A} \ dw(t),\qquad t<0
\end{equation}
It has a unique $\mu^\nu$-stationary solution $\underline v^{\nu,\ep}$;
this process is a strong solution
and has paths in $C((-\infty,0];H^{-\alpha})$ ($\alpha>0$) a.s..
Moreover, the family $\{\underline v^{\nu,\ep}\}_{0<\ep\le 1}$
is tight in $C^{\tilde\beta}([-T,0];H^{-2-\tilde\alpha})$ for
any $0\le \tilde \beta <\frac 12$, $\tilde \alpha>0$ and $T>0$.

Now, we get the existence result.
\begin{theorem}
For any $\nu>0$, there exists a $\mu^\nu$-stationary
process, whose paths solve (a.s.)
equation \eqref{eul} on the time
interval $(-\infty, \infty)$
and  are in $C^\gamma(\mathbb R; H^{-1-\alpha})$
for any $0\le \gamma < 1, \alpha>0$.
\end{theorem}
\proof
Let us fix $\nu>0$. We first construct the solution for $t\ge 0$; then
we can get the result for $t<0$ with the same procedure.

By the tightness result and Prohorov theorem,
the sequence of the laws of $\overline v^{\nu,\ep}$  has a subsequence
$\{\overline v^{\nu,\ep_n}\}_{n=1}^\infty$ weakly
convergent as $n \to \infty$ (with $\ep_n \to 0$)
in $C^\beta([0,T];H^{-2-\alpha})$ to some limit measure.
By a diagonal argument, this holds for any $T$ and therefore the limit
measure $m^\nu$ leaves in  $C^\beta([0,\infty); H^{-2-\alpha})$.
By Skorohod theorem, there
exist a probability space
$(\tilde \Omega^\nu, \tilde {\mathbb F}^\nu, \tilde \po^\nu)$,
a random variable $\tilde v^\nu$ and a sequence $\{\tilde v^{\nu,\ep}\}$
such that law($\tilde v^{\nu,\ep} $)=law($\overline v^{\nu,\ep}$),
law($\tilde v^\nu $)=$m^\nu$ and $\tilde v^{\nu,\ep}$ converges
to $\tilde v^\nu$ a.s. in
$C^\beta([0,\infty); H^{-2-\alpha})$.

We now identify the
equation satisfied by $\tilde v^\nu$. We are going to prove that
  $\tilde \po^\nu$-almost
each path solves \eqref{eul}.
The linear term and the stochastic term, in which appear $\ep$ and
$\sqrt \ep$ respectively, go to zero.
The convergence of the nonlinear term towards
$B(\tilde v^\nu,\tilde v^\nu)$
is proved  by means of the bilinearity
of $B$ and by \eqref{stime-5}.
We have
\[\begin{split}
  \int_0^t
  \|B(\tilde v^{\nu,\ep}(s),&\tilde v^{\nu,\ep}(s))
   -B(\tilde v^{\nu}(s),\tilde v^{\nu}(s))\|_{H^{-5-2\alpha}}ds
  \\&\le
  \int_0^t
   \|B(\tilde v^{\nu,\ep}(s),\tilde v^{\nu,\ep}(s)
         -\tilde v^{\nu}(s))\|_{H^{-5-2\alpha}}ds\\
  &\qquad +\int_0^t
  \|B(\tilde v^{\nu,\ep}(s)-\tilde v^{\nu}(s),\tilde v^{\nu}(s))\|_{H^{-5-2\alpha}}ds
 \\&\le
  ct \|\tilde v^{\nu,\ep}-\tilde v^{\nu}\|_{C([0,t];H^{-2-\alpha})}
     \|\tilde v^{\nu}\|_{C([0,t]; H^{-2-\alpha})} .
\end{split}
\]

The stationarity is inherited from the approximating sequence:
\[
 \tilde{ \mathbb E}^\nu \phi(\tilde v^{\nu,\ep}(t))
 =\int \phi(x)\mu^\nu(dx)
\]
implies
\begin{equation}\label{tildestaz}
\tilde{ \mathbb E}^\nu \phi(\tilde v^{\nu}(t))
 =\int \phi(x)\mu^\nu(dx)
\end{equation}
for any $t\in \mathbb R$ and $\phi \in C_b(H^{-2-\alpha})$.

Finally, from  Proposition \ref{corB} the  right hand side of
\[
 \frac{d \tilde v^{\nu}}{dt}(t)=- B(\tilde v^{\nu}(t),\tilde v^{\nu}(t))
\]
belongs ($\tilde \po^\nu$-a.s.) to
$L^p_{loc}(\mathbb R;H^{-1-\alpha})$ for any $p\in [1,\infty)$ and
  $\alpha>0$.
Hence $\tilde v^{\nu} \in W_{loc}^{1,p}(\mathbb R; H^{-1-\alpha})$;
since $W^{1,p}_{loc}([-T,T])\subset C^\gamma([-T,T])$ for $\gamma < 1-\frac 1p$,
 then $\tilde v^{\nu} \in C^\gamma(\mathbb R; H^{-1-\alpha})$ for $\gamma<1$.
\hfill$\Box$

\medskip
Hence, the paths of the process $\tilde v^\nu$ define a dynamics
for the inviscid shell model \eqref{eul}, having  $\mu^\nu$
as invariant measure.

\end{document}